\documentclass[sigconf]{acmart}
\usepackage{dblfloatfix}

\AtBeginDocument{%
  \providecommand\BibTeX{{%
    \normalfont B\kern-0.5em{\scshape i\kern-0.25em b}\kern-0.8em\TeX}}}

\setcopyright{acmcopyright}
\copyrightyear{2023}
\acmYear{2023}
\setcopyright{rightsretained}
\acmConference[CHI '23 Workshop on Generative AI and HCI]{CHI 2023 Workshop on Generative AI and HCI}{Apr 28, 2023}{Virtual}
\acmBooktitle{CHI 2023 Workshop on Generative AI and HCI), Apr 28, 2023, Virtual}
\acmDOI{10.1145/XXXXXXX.XXXXXXX} 
\begin{document}

\title[LMCanvas: Object-Oriented Interaction for Language Model-Based Writing]{LMCanvas: Object-Oriented Interaction to Personalize Large Language Model-Powered Writing Environments}

\author{Tae Soo Kim}
\email{taesoo.kim@kaist.ac.kr}
\affiliation{%
  \institution{School of Computing, KAIST}
  \city{Daejeon}
  \country{Republic of Korea}
}

\author{Arghya Sarkar}
\email{arghya@nyu.edu}
\affiliation{%
  \institution{New York University}
  \city{New York}
  \country{USA}
}

\author{Yoonjoo Lee}
\email{yoonjoo.lee@kaist.ac.kr}
\affiliation{%
  \institution{School of Computing, KAIST}
  \city{Daejeon}
  \country{Republic of Korea}
}

\author{Minsuk Chang}
\authornote{Minsuk is now at Google.}
\email{minsuk.chang@navercorp.com}
\affiliation{%
  \institution{Naver AI Lab}
  \city{Seongnam}
  \country{Republic of Korea}
}

\author{Juho Kim}
\email{juhokim@kaist.ac.kr}
\affiliation{%
  \institution{School of Computing, KAIST}
  \city{Daejeon}
  \country{Republic of Korea}
}

\renewcommand{\shortauthors}{Kim et al.}

\begin{abstract}

Large language models (LLMs) can enhance writing by automating or supporting specific tasks in writers' workflows (e.g., paraphrasing, creating analogies).
Leveraging this capability, a collection of interfaces have been developed that provide LLM-powered tools for specific writing tasks. However, these interfaces provide limited support for writers to create personal tools for their own unique tasks, and may not comprehensively fulfill a writer’s needs---requiring them to continuously switch between interfaces during writing.
In this work, we envision LMCanvas, an interface that enables writers to create their own LLM-powered writing tools and arrange their personal writing environment by interacting with ``blocks’’ in a canvas. 
In this interface, users can create text blocks to encapsulate writing and LLM prompts, model blocks for model parameter configurations, and connect these to create pipeline blocks that output generations.
In this workshop paper, we discuss the design for LMCanvas and our plans to develop this concept.

\end{abstract}



\maketitle

\section{Introduction}

The advent of large language models (LLMs)---e.g., GPT-3~\cite{brown2020language}, GPT-NeoX~\cite{gpt-neox-20b}, Jurassic-1~\cite{J1WhitePaper}, LaMDA~\cite{lamda}---has transformed the writing process.
Instead of manually drafting passages of text, writers can now hand over this effort to these models and almost instantly generate passages from an initial sentence or phrase.
Beyond their generative capabilities, LLMs demonstrate significant few-shot and zero-shot performance~\cite{brown2020language} meaning that they are able to perform previously unseen tasks with only an instruction and/or a couple of examples---i.e., a prompt.
By leveraging this ability of LLMs, writers can potentially automate or augment specific tasks in their workflows by using adequate prompts and, thus, further facilitate the writing process.
For instance, based only on prompt examples provided for GPT-3~\cite{examplesopenai}, writers can use LLMs to correct grammar, create an outline, produce analogies, or even change the point-of-view of a scene.

To seize the opportunity presented by LLMs, an assortment of products and interfaces have been created that leverage these models to provide writers with specific tools that automate steps in their writing workflows. 
For example, tools such as WordTune~\cite{wordtune} and NotionAI~\cite{notionai} provide editing buttons that the user can click after selecting text to automatically rewrite it, change its tone, summarize it, elaborate on it, etc.
Additionally, a variety of LLM-powered copywriting tools~\cite{jasper, copyai} have also been created that provide writers with a variety of template forms that they can fill to generate specific types of writing (e.g., video description or script, blog introduction, article headline).
Similarly in academia, various interfaces have been designed to leverage LLMs to support specific tasks: generate various forms of figurative language~\cite{chakrabarty2022help}, summarize a writer's writing~\cite{dang2022beyond}, brainstorm and combine ideas~\cite{di2022idea}, or propagate writing edits across a story~\cite{lee2022interactive}.

\begin{figure}[hb]
    \centering
    \includegraphics[width=1.0\columnwidth]{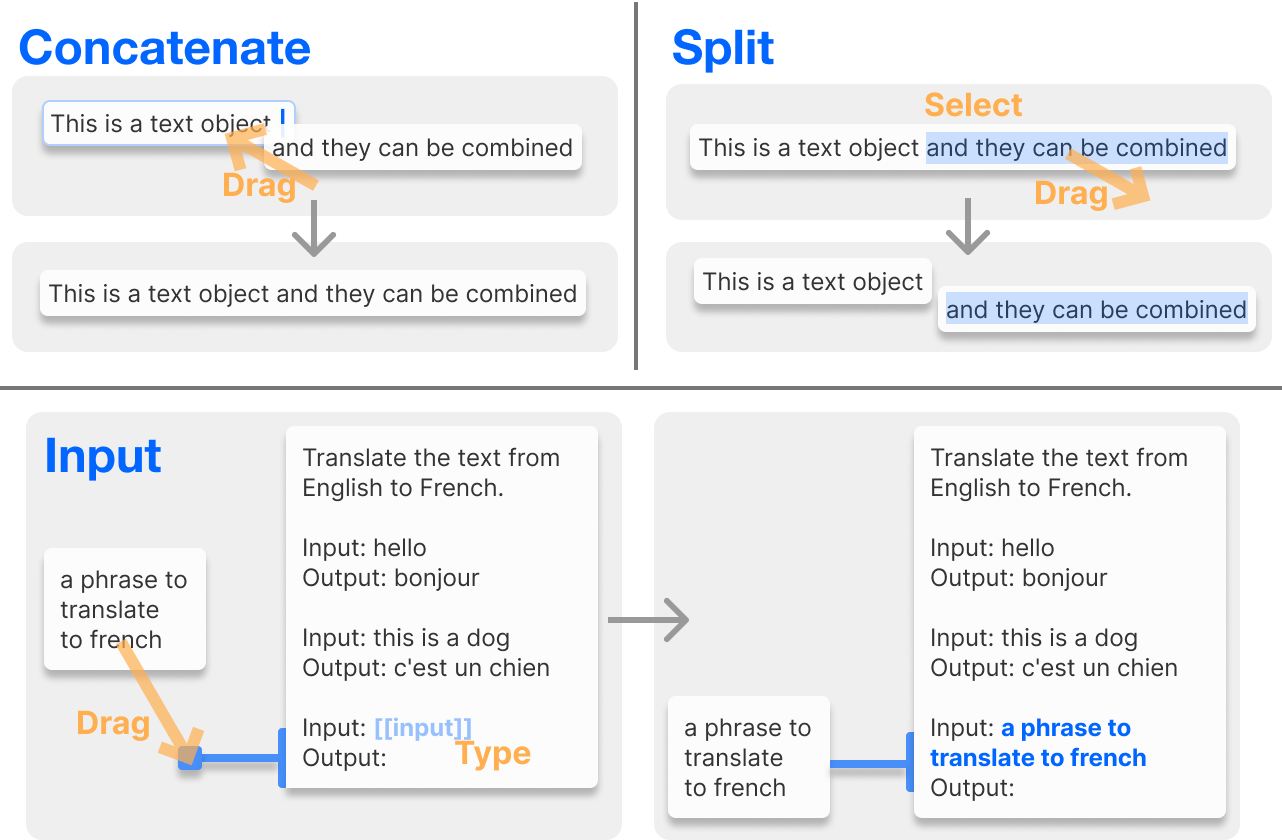}
    \caption{Illustrations for the \textbf{concatenate}, \textbf{split}, and \textbf{input} interactions supported in text blocks.}
    \label{fig:textblock}
\end{figure}

While the proliferation of these LLM-driven tools means that various writing tasks can now be supported, the individual needs and challenges of writers might not be fulfilled by these tools.
Due to their type of writing, their fluency with a language, or other factors such as their style and workflow, a writer may have specific needs and challenges during their writing process.
However, while existing interfaces provide a general set of tools, they provide limited or no support for the writer to create their own tools to support their unique tasks.
Further, an interface may not provide a comprehensive set of tools that supports all of the writer's tasks and, thus, the writer may need to constantly switch between multiple interfaces to support their workflows.
As a result, the writer needs to scatter and adapt their writing workflow across a variety of interfaces.

In this work, we envision a canvas-based interface that enables writers to create their own personalized LLM-driven tools and configure them into one cohesive writing environment.
Inspired by object-oriented interaction~\cite{xia2016object, ciolfi2016beyond, xia2018dataink, xia2017collection, han2022passages, beaudouin2000instrumental} and block-based programming~\cite{resnick2009scratch},  we present the design for \textbf{LMCanvas}, an interface that enables users to interact with text and model blocks to flexibly create and arrange LLM-powered tools.
Through the interface, users can create text blocks to encapsulate both their writing and LLM prompts, keep drafts as separate blocks, and organize them in the canvas. 
By connecting text blocks to model blocks (i.e., blocks that represent a set of model configurations), users can create LM pipelines, tools, that generate outputs as text blocks based on the input text and model block.
After creating a set of tools, the user can flexibly arrange them in the canvas to create a writing environment customized to their needs and preferences. 
In this workshop paper, we discuss our design of the envisioned LMCanvas and future work to develop this concept. 

\section{LMCanvas: Design Concept}

In our envisioned interface, \textbf{LMCanvas}, writers can create four types of objects in an infinite canvas: \textit{text blocks}, \textit{model blocks}, and \textit{pipeline blocks}.
All of these blocks can be flexibly moved, copy-pasted, deleted, and connected to each other.
Below, we detail the specific interactions that we aim to support for each type of block.

\subsubsection{Text Blocks}

In the canvas, the writer can create text blocks, which are objects that the writer can type text into and edit.
When writing with LLMs, text can represent different types of content: actual writing, prompts, examples for prompts, generated outputs, etc.
By compartmentalizing text into modular blocks, our interface allows the writer to flexibly organize and structure these different forms of text in their writing environment.
For example, the writer can use a text block as their main text editor, maintain text blocks on the side containing alternative versions for certain paragraphs, and keep a text block with a prompt template to reuse when creating LLM-powered tools.

\begin{figure}[b]
    \centering
    \includegraphics[width=0.8\columnwidth]{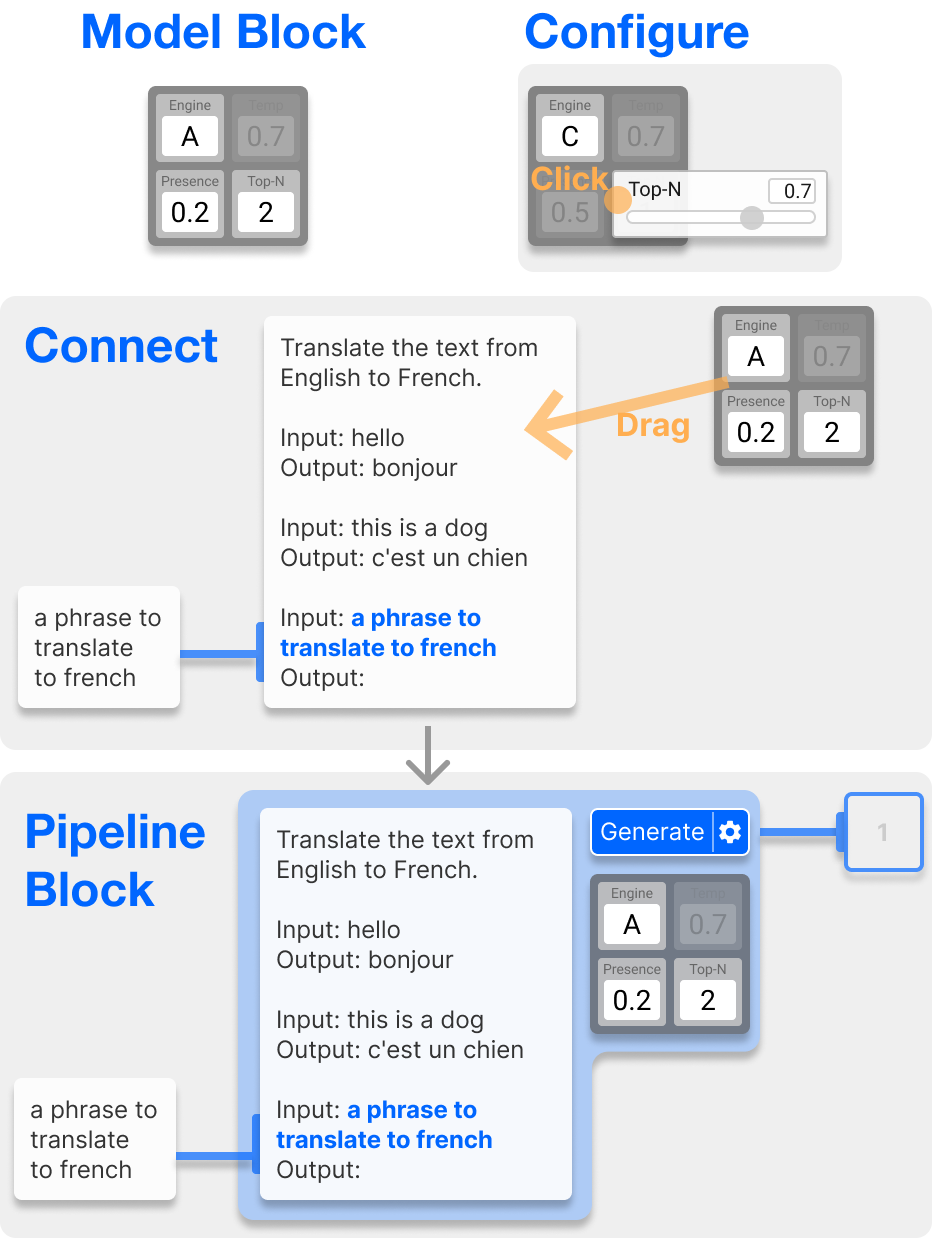}
    \caption{Model blocks represent a set of parameter configurations that the writer can configure, copy, and reuse. By connecting model blocks to text blocks, writers can create pipeline blocks that allow them to generate outputs based on the nested text and parameters.}
    \label{fig:modelblock}
\end{figure}

To support flexible use of text blocks, we design the following interactions specific to these blocks. 
\textbf{Resizing} allows the writer to change the format of text blocks for different types of usage (e.g., a larger block for a text editor) or to decrease their size to decrease clutter in the screen.
When the writer decides that they do not have to keep two text blocks separate anymore (e.g., decided on the final versions for the first two verses of their poem), they can \textbf{concatenate} these blocks by drag-and-dropping one text block into the other.
Alternatively, if the writer needs to modularize or separate certain parts of a text (e.g., to only draft one part of a paragraph), they can \textbf{split} off text by selecting it and dragging it outwards---creating a new text block.
To allow writers to create reusable LLM-powered tools, the interface allows the user to create text blocks to which  they can \textbf{input} other text blocks.
Specifically, the user types the ``\verb|[[input]]|'' command in a text block to create an ``input prong''. Then they can attach other blocks into this prong to replace the ``\verb|[[input]]|'' command with the content of the attached text block.
Finally, as writers may want writing support tools to act on selected text (e.g., generate metaphor for selected phrase or edit selected text to be shorter), the interface also allows users to create \textbf{select} blocks by typing the ``\verb|[[select]]|'' command in a text block.
The content of these blocks are replaced by any text that the user selects in the canvas.

\begin{figure*}[!ht]
    \centering
    \includegraphics[width=1.0\textwidth]{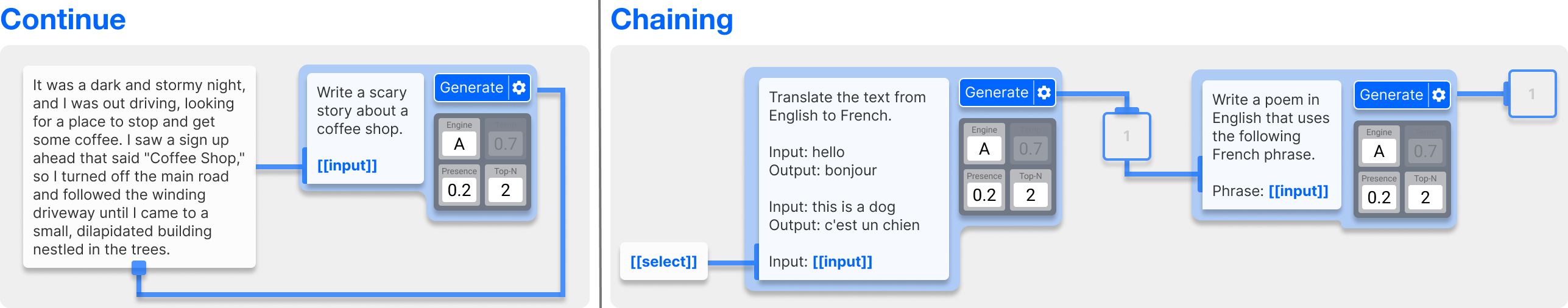}
    \caption{The output container of pipeline blocks can be connected to text blocks to add generations as continuations, or to the input prongs of text blocks to chain pipelines.}
    \label{fig:pipelineblocks}
\end{figure*}

\subsubsection{Model Blocks}

LLMs possess various parameters that control the generation process. 
For example, the temperature parameter determines the probability of the model  generating more out-of-distribution or improbable text. 
Prior work has demonstrated that, when writing with LLMs, different configurations of these parameters can satisfy different user needs~\cite{lee2022coauthor}.
Thus, to support writers to set, test, and reuse parameter configurations, LMCanvas allows users to create multiple model blocks with different combinations of parameters (\autoref{fig:modelblock}).
These blocks represents an instance of parameter configurations that the writer can re\textbf{configure} by clicking on a parameter and using the displayed widgets to change its value. 

\subsubsection{Pipeline Blocks}

To generate text with the LLM, the user can \textbf{connect} a text block to a model block to create a pipeline block (\autoref{fig:modelblock}).
When a writer clicks on ``generate'' in a pipeline block, the interface uses the nested text block as input and the model block as the parameter configurations to generate an output, which is presented as a text block.
To test multiple inputs and parameter configurations, the writer can also \textbf{expand} a pipeline block by adding additional text and model blocks.
In this case, when the pipeline block generates, it produces a generation for each pairing of text and model blocks inside the pipeline block.

By default, each time a pipeline block generates, it adds the generation as a text block in the output container---the box containing ``1'' that prongs out of the pipeline block as seen in \autoref{fig:modelblock}.
However, through drag-and-drop, the writer can connect this output container to (1) a text block to add the generations from the pipeline as \textbf{continuations} to that text block (left in \autoref{fig:pipelineblocks}), or (2) an input prong to \textbf{chain} multiple pipeline blocks together and create more complex tools~\cite{wu2021aichains} (right in \autoref{fig:pipelineblocks}).
Additionally, if the writer connects the output of a pipeline to a \textbf{select} block, the interface replaces any text selected across all text blocks in the canvas with the generation produced by the pipeline.
These various forms of connecting pipeline blocks can enable the writer to create a variety of tools from the same basic blocks.
\section{Future Work}

In this workshop paper, we outlined the foundational objects and interactions that we aim to support in our envisioned LMCanvas.
Our goal with this interface is to enable writers to more effectively leverage LLMs by personalizing their use to fit their unique workflows, needs, and challenges.
At the current stage of the project, we have developed an initial prototype that supports the three objects presented in this paper and their basic interactions.
With this prototype, we are planning to conduct formative studies to understand the various tools that writers can create with LMCanvas, the benefits and drawbacks of the interfaces, and additional blocks and interactions that writers might need.
Based on the findings, we plan to improve and expand on the concept.
Beyond the directions for improvement to be distilled from the formative study, there are additional future directions that we plan to pursue with LMCanvas.

First, an additional benefit of representing writing as text blocks is that this can enable the interface to maintain a separate history for each text block.
With this modularized history, writers can check on and revert changes for only specific parts of their writing, and they can also reflect back on their generation attempts by seeing what inputs and parameter configurations were previously used.
We are planning to implement this modularized history for text blocks and to enable users to interact with it---e.g., dragging the text input that generated a text block out from the history and into the canvas.

Second, in future versions of LMCanvas, we aim to support various types of output containers for pipeline blocks. 
Currently, the prototype only supports containers that keep generated text blocks as a list.
However, when dealing with a large quantity of generated outputs, writers may need alternative methods to look at and explore generated outputs.
For example, generations could be encoded in a scatterplot~\cite{matejka2018dreamlens} to enable the writer to visualize the output space. 

Finally, identifying effective prompts (i.e., prompt engineering) is a major hurdle in leveraging LLMs. 
While tools have been designed to facilitate this in well-defined tasks where there is a ``ground-truth''~\cite{strobelt2022promptide}, there is limited work that investigated how to support prompt engineering in open-ended and more creative tasks.
Through our initial versions of LMCanvas, we aim to investigate mechanisms to facilitate prompt engineering in open-ended writing and to incorporate these into the interface.
For example, the interface could allow writers to drag-and-drop text blocks into pipeline blocks as positive or negative examples, and leverage these in the back-end to produce outputs more aligned to the writers' preferences.

\begin{acks}
This work was supported by KAIST-NAVER Hypercreative AI Center.
\end{acks}

\bibliographystyle{ACM-Reference-Format}
\bibliography{bibliography}


\begin{thebibliography}{24}


\ifx \showCODEN    \undefined \def \showCODEN     #1{\unskip}     \fi
\ifx \showDOI      \undefined \def \showDOI       #1{#1}\fi
\ifx \showISBNx    \undefined \def \showISBNx     #1{\unskip}     \fi
\ifx \showISBNxiii \undefined \def \showISBNxiii  #1{\unskip}     \fi
\ifx \showISSN     \undefined \def \showISSN      #1{\unskip}     \fi
\ifx \showLCCN     \undefined \def \showLCCN      #1{\unskip}     \fi
\ifx \shownote     \undefined \def \shownote      #1{#1}          \fi
\ifx \showarticletitle \undefined \def \showarticletitle #1{#1}   \fi
\ifx \showURL      \undefined \def \showURL       {\relax}        \fi
\providecommand\bibfield[2]{#2}
\providecommand\bibinfo[2]{#2}
\providecommand\natexlab[1]{#1}
\providecommand\showeprint[2][]{arXiv:#2}

\bibitem[AI(2022)]%
        {jasper}
\bibfield{author}{\bibinfo{person}{Jasper AI}.}
  \bibinfo{year}{2022}\natexlab{}.
\newblock \bibinfo{booktitle}{\emph{Jasper - AI Copywriter | AI Content
  Generator for Teams}}.
\newblock
\urldef\tempurl%
\url{https://www.jasper.ai/}
\showURL{%
Retrieved Feb 23, 2023 from \tempurl}


\bibitem[Beaudouin-Lafon(2000)]%
        {beaudouin2000instrumental}
\bibfield{author}{\bibinfo{person}{Michel Beaudouin-Lafon}.}
  \bibinfo{year}{2000}\natexlab{}.
\newblock \showarticletitle{Instrumental Interaction: An Interaction Model for
  Designing Post-WIMP User Interfaces}. In
  \bibinfo{booktitle}{\emph{Proceedings of the SIGCHI Conference on Human
  Factors in Computing Systems}} (The Hague, The Netherlands)
  \emph{(\bibinfo{series}{CHI '00})}. \bibinfo{publisher}{Association for
  Computing Machinery}, \bibinfo{address}{New York, NY, USA},
  \bibinfo{pages}{446–453}.
\newblock
\showISBNx{1581132166}
\urldef\tempurl%
\url{https://doi.org/10.1145/332040.332473}
\showDOI{\tempurl}


\bibitem[Black et~al\mbox{.}(2022)]%
        {gpt-neox-20b}
\bibfield{author}{\bibinfo{person}{Sid Black}, \bibinfo{person}{Stella
  Biderman}, \bibinfo{person}{Eric Hallahan}, \bibinfo{person}{Quentin
  Anthony}, \bibinfo{person}{Leo Gao}, \bibinfo{person}{Laurence Golding},
  \bibinfo{person}{Horace He}, \bibinfo{person}{Connor Leahy},
  \bibinfo{person}{Kyle McDonell}, \bibinfo{person}{Jason Phang},
  \bibinfo{person}{Michael Pieler}, \bibinfo{person}{USVSN~Sai Prashanth},
  \bibinfo{person}{Shivanshu Purohit}, \bibinfo{person}{Laria Reynolds},
  \bibinfo{person}{Jonathan Tow}, \bibinfo{person}{Ben Wang}, {and}
  \bibinfo{person}{Samuel Weinbach}.} \bibinfo{year}{2022}\natexlab{}.
\newblock \showarticletitle{{GPT-NeoX-20B}: An Open-Source Autoregressive
  Language Model}.
\newblock  (\bibinfo{year}{2022}).
\newblock


\bibitem[Brown et~al\mbox{.}(2020)]%
        {brown2020language}
\bibfield{author}{\bibinfo{person}{Tom~B. Brown}, \bibinfo{person}{Benjamin
  Mann}, \bibinfo{person}{Nick Ryder}, \bibinfo{person}{Melanie Subbiah},
  \bibinfo{person}{Jared Kaplan}, \bibinfo{person}{Prafulla Dhariwal},
  \bibinfo{person}{Arvind Neelakantan}, \bibinfo{person}{Pranav Shyam},
  \bibinfo{person}{Girish Sastry}, \bibinfo{person}{Amanda Askell},
  \bibinfo{person}{Sandhini Agarwal}, \bibinfo{person}{Ariel Herbert-Voss},
  \bibinfo{person}{Gretchen Krueger}, \bibinfo{person}{Tom Henighan},
  \bibinfo{person}{Rewon Child}, \bibinfo{person}{Aditya Ramesh},
  \bibinfo{person}{Daniel~M. Ziegler}, \bibinfo{person}{Jeffrey Wu},
  \bibinfo{person}{Clemens Winter}, \bibinfo{person}{Christopher Hesse},
  \bibinfo{person}{Mark Chen}, \bibinfo{person}{Eric Sigler},
  \bibinfo{person}{Mateusz Litwin}, \bibinfo{person}{Scott Gray},
  \bibinfo{person}{Benjamin Chess}, \bibinfo{person}{Jack Clark},
  \bibinfo{person}{Christopher Berner}, \bibinfo{person}{Sam McCandlish},
  \bibinfo{person}{Alec Radford}, \bibinfo{person}{Ilya Sutskever}, {and}
  \bibinfo{person}{Dario Amodei}.} \bibinfo{year}{2020}\natexlab{}.
\newblock \bibinfo{title}{Language Models are Few-Shot Learners}.
\newblock
\newblock
\showeprint[arxiv]{2005.14165}~[cs.CL]


\bibitem[Chakrabarty et~al\mbox{.}(2022)]%
        {chakrabarty2022help}
\bibfield{author}{\bibinfo{person}{Tuhin Chakrabarty}, \bibinfo{person}{Vishakh
  Padmakumar}, {and} \bibinfo{person}{He He}.} \bibinfo{year}{2022}\natexlab{}.
\newblock \showarticletitle{Help me write a poem: Instruction Tuning as a
  Vehicle for Collaborative Poetry Writing}.
\newblock \bibinfo{journal}{\emph{arXiv preprint arXiv:2210.13669}}
  (\bibinfo{year}{2022}).
\newblock


\bibitem[Ciolfi~Felice et~al\mbox{.}(2016)]%
        {ciolfi2016beyond}
\bibfield{author}{\bibinfo{person}{Marianela Ciolfi~Felice},
  \bibinfo{person}{Nolwenn Maudet}, \bibinfo{person}{Wendy~E. Mackay}, {and}
  \bibinfo{person}{Michel Beaudouin-Lafon}.} \bibinfo{year}{2016}\natexlab{}.
\newblock \showarticletitle{Beyond Snapping: Persistent, Tweakable Alignment
  and Distribution with StickyLines}. In \bibinfo{booktitle}{\emph{Proceedings
  of the 29th Annual Symposium on User Interface Software and Technology}}
  (Tokyo, Japan) \emph{(\bibinfo{series}{UIST '16})}.
  \bibinfo{publisher}{Association for Computing Machinery},
  \bibinfo{address}{New York, NY, USA}, \bibinfo{pages}{133–144}.
\newblock
\showISBNx{9781450341899}
\urldef\tempurl%
\url{https://doi.org/10.1145/2984511.2984577}
\showDOI{\tempurl}


\bibitem[CopyAI(2022)]%
        {copyai}
\bibfield{author}{\bibinfo{person}{CopyAI}.} \bibinfo{year}{2022}\natexlab{}.
\newblock \bibinfo{booktitle}{\emph{Copy.ai: Write better marketing copy and
  content with AI}}.
\newblock
\urldef\tempurl%
\url{https://www.copy.ai/}
\showURL{%
Retrieved Feb 23, 2023 from \tempurl}


\bibitem[Dang et~al\mbox{.}(2022)]%
        {dang2022beyond}
\bibfield{author}{\bibinfo{person}{Hai Dang}, \bibinfo{person}{Karim
  Benharrak}, \bibinfo{person}{Florian Lehmann}, {and} \bibinfo{person}{Daniel
  Buschek}.} \bibinfo{year}{2022}\natexlab{}.
\newblock \showarticletitle{Beyond Text Generation: Supporting Writers with
  Continuous Automatic Text Summaries}. In
  \bibinfo{booktitle}{\emph{Proceedings of the 35th Annual ACM Symposium on
  User Interface Software and Technology}}. \bibinfo{pages}{1--13}.
\newblock


\bibitem[Di~Fede et~al\mbox{.}(2022)]%
        {di2022idea}
\bibfield{author}{\bibinfo{person}{Giulia Di~Fede}, \bibinfo{person}{Davide
  Rocchesso}, \bibinfo{person}{Steven~P Dow}, {and} \bibinfo{person}{Salvatore
  Andolina}.} \bibinfo{year}{2022}\natexlab{}.
\newblock \showarticletitle{The Idea Machine: LLM-based Expansion, Rewriting,
  Combination, and Suggestion of Ideas}. In
  \bibinfo{booktitle}{\emph{Creativity and Cognition}}.
  \bibinfo{pages}{623--627}.
\newblock


\bibitem[Han et~al\mbox{.}(2022)]%
        {han2022passages}
\bibfield{author}{\bibinfo{person}{Han~L. Han}, \bibinfo{person}{Junhang Yu},
  \bibinfo{person}{Raphael Bournet}, \bibinfo{person}{Alexandre Ciorascu},
  \bibinfo{person}{Wendy~E. Mackay}, {and} \bibinfo{person}{Michel
  Beaudouin-Lafon}.} \bibinfo{year}{2022}\natexlab{}.
\newblock \showarticletitle{Passages: Interacting with Text Across Documents}.
  In \bibinfo{booktitle}{\emph{Proceedings of the 2022 CHI Conference on Human
  Factors in Computing Systems}} (New Orleans, LA, USA)
  \emph{(\bibinfo{series}{CHI '22})}. \bibinfo{publisher}{Association for
  Computing Machinery}, \bibinfo{address}{New York, NY, USA}, Article
  \bibinfo{articleno}{338}, \bibinfo{numpages}{17}~pages.
\newblock
\showISBNx{9781450391573}
\urldef\tempurl%
\url{https://doi.org/10.1145/3491102.3502052}
\showDOI{\tempurl}


\bibitem[Labs(2022a)]%
        {wordtune}
\bibfield{author}{\bibinfo{person}{AI21 Labs}.}
  \bibinfo{year}{2022}\natexlab{a}.
\newblock \bibinfo{booktitle}{\emph{WordTune | Your personal writing assistant
  and editor}}.
\newblock
\urldef\tempurl%
\url{https://www.wordtune.com/}
\showURL{%
Retrieved Feb 23, 2023 from \tempurl}


\bibitem[Labs(2022b)]%
        {notionai}
\bibfield{author}{\bibinfo{person}{Notion Labs}.}
  \bibinfo{year}{2022}\natexlab{b}.
\newblock \bibinfo{booktitle}{\emph{Notion AI}}.
\newblock
\urldef\tempurl%
\url{https://www.notion.so/product/ai}
\showURL{%
Retrieved Feb 23, 2023 from \tempurl}


\bibitem[Lee et~al\mbox{.}(2022b)]%
        {lee2022coauthor}
\bibfield{author}{\bibinfo{person}{Mina Lee}, \bibinfo{person}{Percy Liang},
  {and} \bibinfo{person}{Qian Yang}.} \bibinfo{year}{2022}\natexlab{b}.
\newblock \showarticletitle{CoAuthor: Designing a Human-AI Collaborative
  Writing Dataset for Exploring Language Model Capabilities}.
\newblock \bibinfo{journal}{\emph{CoRR}}  \bibinfo{volume}{abs/2201.06796}
  (\bibinfo{year}{2022}).
\newblock
\showeprint[arXiv]{2201.06796}
\urldef\tempurl%
\url{https://arxiv.org/abs/2201.06796}
\showURL{%
\tempurl}


\bibitem[Lee et~al\mbox{.}(2022a)]%
        {lee2022interactive}
\bibfield{author}{\bibinfo{person}{Yoonjoo Lee}, \bibinfo{person}{Tae~Soo Kim},
  \bibinfo{person}{Minsuk Chang}, {and} \bibinfo{person}{Juho Kim}.}
  \bibinfo{year}{2022}\natexlab{a}.
\newblock \showarticletitle{Interactive Children’s Story Rewriting Through
  Parent-Children Interaction}. In \bibinfo{booktitle}{\emph{Proceedings of the
  First Workshop on Intelligent and Interactive Writing Assistants (In2Writing
  2022)}}. \bibinfo{pages}{62--71}.
\newblock


\bibitem[Lieber et~al\mbox{.}(2021)]%
        {J1WhitePaper}
\bibfield{author}{\bibinfo{person}{Opher Lieber}, \bibinfo{person}{Or Sharir},
  \bibinfo{person}{Barak Lenz}, {and} \bibinfo{person}{Yoav Shoham}.}
  \bibinfo{year}{2021}\natexlab{}.
\newblock \bibinfo{booktitle}{\emph{Jurassic-1: Technical Details And
  Evaluation}}.
\newblock \bibinfo{type}{{T}echnical {R}eport}. \bibinfo{institution}{AI21
  Labs}.
\newblock


\bibitem[Matejka et~al\mbox{.}(2018)]%
        {matejka2018dreamlens}
\bibfield{author}{\bibinfo{person}{Justin Matejka}, \bibinfo{person}{Michael
  Glueck}, \bibinfo{person}{Erin Bradner}, \bibinfo{person}{Ali Hashemi},
  \bibinfo{person}{Tovi Grossman}, {and} \bibinfo{person}{George Fitzmaurice}.}
  \bibinfo{year}{2018}\natexlab{}.
\newblock \bibinfo{booktitle}{\emph{Dream Lens: Exploration and Visualization
  of Large-Scale Generative Design Datasets}}.
\newblock \bibinfo{publisher}{Association for Computing Machinery},
  \bibinfo{address}{New York, NY, USA}, \bibinfo{pages}{1–12}.
\newblock
\showISBNx{9781450356206}
\urldef\tempurl%
\url{https://doi.org/10.1145/3173574.3173943}
\showURL{%
\tempurl}


\bibitem[OpenAI(2022)]%
        {examplesopenai}
\bibfield{author}{\bibinfo{person}{OpenAI}.} \bibinfo{year}{2022}\natexlab{}.
\newblock \bibinfo{booktitle}{\emph{Examples - OpenAI API}}.
\newblock
\urldef\tempurl%
\url{https://platform.openai.com/examples/}
\showURL{%
Retrieved Feb 23, 2023 from \tempurl}


\bibitem[Resnick et~al\mbox{.}(2009)]%
        {resnick2009scratch}
\bibfield{author}{\bibinfo{person}{Mitchel Resnick}, \bibinfo{person}{John
  Maloney}, \bibinfo{person}{Andr{\'e}s Monroy-Hern{\'a}ndez},
  \bibinfo{person}{Natalie Rusk}, \bibinfo{person}{Evelyn Eastmond},
  \bibinfo{person}{Karen Brennan}, \bibinfo{person}{Amon Millner},
  \bibinfo{person}{Eric Rosenbaum}, \bibinfo{person}{Jay Silver},
  \bibinfo{person}{Brian Silverman}, {et~al\mbox{.}}}
  \bibinfo{year}{2009}\natexlab{}.
\newblock \showarticletitle{Scratch: programming for all}.
\newblock \bibinfo{journal}{\emph{Commun. ACM}} \bibinfo{volume}{52},
  \bibinfo{number}{11} (\bibinfo{year}{2009}), \bibinfo{pages}{60--67}.
\newblock


\bibitem[Strobelt et~al\mbox{.}(2022)]%
        {strobelt2022promptide}
\bibfield{author}{\bibinfo{person}{Hendrik Strobelt}, \bibinfo{person}{Albert
  Webson}, \bibinfo{person}{Victor Sanh}, \bibinfo{person}{Benjamin Hoover},
  \bibinfo{person}{Johanna Beyer}, \bibinfo{person}{Hanspeter Pfister}, {and}
  \bibinfo{person}{Alexander~M. Rush}.} \bibinfo{year}{2022}\natexlab{}.
\newblock \bibinfo{title}{Interactive and Visual Prompt Engineering for Ad-hoc
  Task Adaptation with Large Language Models}.
\newblock
\newblock
\urldef\tempurl%
\url{https://doi.org/10.48550/ARXIV.2208.07852}
\showDOI{\tempurl}


\bibitem[Thoppilan et~al\mbox{.}(2022)]%
        {lamda}
\bibfield{author}{\bibinfo{person}{Romal Thoppilan}, \bibinfo{person}{Daniel~De
  Freitas}, \bibinfo{person}{Jamie Hall}, \bibinfo{person}{Noam Shazeer},
  \bibinfo{person}{Apoorv Kulshreshtha}, \bibinfo{person}{Heng{-}Tze Cheng},
  \bibinfo{person}{Alicia Jin}, \bibinfo{person}{Taylor Bos},
  \bibinfo{person}{Leslie Baker}, \bibinfo{person}{Yu Du},
  \bibinfo{person}{YaGuang Li}, \bibinfo{person}{Hongrae Lee},
  \bibinfo{person}{Huaixiu~Steven Zheng}, \bibinfo{person}{Amin Ghafouri},
  \bibinfo{person}{Marcelo Menegali}, \bibinfo{person}{Yanping Huang},
  \bibinfo{person}{Maxim Krikun}, \bibinfo{person}{Dmitry Lepikhin},
  \bibinfo{person}{James Qin}, \bibinfo{person}{Dehao Chen},
  \bibinfo{person}{Yuanzhong Xu}, \bibinfo{person}{Zhifeng Chen},
  \bibinfo{person}{Adam Roberts}, \bibinfo{person}{Maarten Bosma},
  \bibinfo{person}{Yanqi Zhou}, \bibinfo{person}{Chung{-}Ching Chang},
  \bibinfo{person}{Igor Krivokon}, \bibinfo{person}{Will Rusch},
  \bibinfo{person}{Marc Pickett}, \bibinfo{person}{Kathleen~S.
  Meier{-}Hellstern}, \bibinfo{person}{Meredith~Ringel Morris},
  \bibinfo{person}{Tulsee Doshi}, \bibinfo{person}{Renelito~Delos Santos},
  \bibinfo{person}{Toju Duke}, \bibinfo{person}{Johnny Soraker},
  \bibinfo{person}{Ben Zevenbergen}, \bibinfo{person}{Vinodkumar Prabhakaran},
  \bibinfo{person}{Mark Diaz}, \bibinfo{person}{Ben Hutchinson},
  \bibinfo{person}{Kristen Olson}, \bibinfo{person}{Alejandra Molina},
  \bibinfo{person}{Erin Hoffman{-}John}, \bibinfo{person}{Josh Lee},
  \bibinfo{person}{Lora Aroyo}, \bibinfo{person}{Ravi Rajakumar},
  \bibinfo{person}{Alena Butryna}, \bibinfo{person}{Matthew Lamm},
  \bibinfo{person}{Viktoriya Kuzmina}, \bibinfo{person}{Joe Fenton},
  \bibinfo{person}{Aaron Cohen}, \bibinfo{person}{Rachel Bernstein},
  \bibinfo{person}{Ray Kurzweil}, \bibinfo{person}{Blaise Aguera{-}Arcas},
  \bibinfo{person}{Claire Cui}, \bibinfo{person}{Marian Croak},
  \bibinfo{person}{Ed Chi}, {and} \bibinfo{person}{Quoc Le}.}
  \bibinfo{year}{2022}\natexlab{}.
\newblock \showarticletitle{LaMDA: Language Models for Dialog Applications}.
\newblock \bibinfo{journal}{\emph{CoRR}}  \bibinfo{volume}{abs/2201.08239}
  (\bibinfo{year}{2022}).
\newblock
\showeprint[arXiv]{2201.08239}
\urldef\tempurl%
\url{https://arxiv.org/abs/2201.08239}
\showURL{%
\tempurl}


\bibitem[Wu et~al\mbox{.}(2021)]%
        {wu2021aichains}
\bibfield{author}{\bibinfo{person}{Tongshuang Wu}, \bibinfo{person}{Michael
  Terry}, {and} \bibinfo{person}{Carrie~J. Cai}.}
  \bibinfo{year}{2021}\natexlab{}.
\newblock \showarticletitle{{AI} Chains: Transparent and Controllable Human-AI
  Interaction by Chaining Large Language Model Prompts}.
\newblock \bibinfo{journal}{\emph{CoRR}}  \bibinfo{volume}{abs/2110.01691}
  (\bibinfo{year}{2021}).
\newblock
\showeprint[arXiv]{2110.01691}
\urldef\tempurl%
\url{https://arxiv.org/abs/2110.01691}
\showURL{%
\tempurl}


\bibitem[Xia et~al\mbox{.}(2016)]%
        {xia2016object}
\bibfield{author}{\bibinfo{person}{Haijun Xia}, \bibinfo{person}{Bruno Araujo},
  \bibinfo{person}{Tovi Grossman}, {and} \bibinfo{person}{Daniel Wigdor}.}
  \bibinfo{year}{2016}\natexlab{}.
\newblock \showarticletitle{Object-Oriented Drawing}. In
  \bibinfo{booktitle}{\emph{Proceedings of the 2016 CHI Conference on Human
  Factors in Computing Systems}} (San Jose, California, USA)
  \emph{(\bibinfo{series}{CHI '16})}. \bibinfo{publisher}{Association for
  Computing Machinery}, \bibinfo{address}{New York, NY, USA},
  \bibinfo{pages}{4610–4621}.
\newblock
\showISBNx{9781450333627}
\urldef\tempurl%
\url{https://doi.org/10.1145/2858036.2858075}
\showDOI{\tempurl}


\bibitem[Xia et~al\mbox{.}(2017)]%
        {xia2017collection}
\bibfield{author}{\bibinfo{person}{Haijun Xia}, \bibinfo{person}{Bruno Araujo},
  {and} \bibinfo{person}{Daniel Wigdor}.} \bibinfo{year}{2017}\natexlab{}.
\newblock \showarticletitle{Collection Objects: Enabling Fluid Formation and
  Manipulation of Aggregate Selections}. In
  \bibinfo{booktitle}{\emph{Proceedings of the 2017 CHI Conference on Human
  Factors in Computing Systems}} (Denver, Colorado, USA)
  \emph{(\bibinfo{series}{CHI '17})}. \bibinfo{publisher}{Association for
  Computing Machinery}, \bibinfo{address}{New York, NY, USA},
  \bibinfo{pages}{5592–5604}.
\newblock
\showISBNx{9781450346559}
\urldef\tempurl%
\url{https://doi.org/10.1145/3025453.3025554}
\showDOI{\tempurl}


\bibitem[Xia et~al\mbox{.}(2018)]%
        {xia2018dataink}
\bibfield{author}{\bibinfo{person}{Haijun Xia}, \bibinfo{person}{Nathalie
  Henry~Riche}, \bibinfo{person}{Fanny Chevalier}, \bibinfo{person}{Bruno
  De~Araujo}, {and} \bibinfo{person}{Daniel Wigdor}.}
  \bibinfo{year}{2018}\natexlab{}.
\newblock \bibinfo{booktitle}{\emph{DataInk: Direct and Creative Data-Oriented
  Drawing}}.
\newblock \bibinfo{publisher}{Association for Computing Machinery},
  \bibinfo{address}{New York, NY, USA}, \bibinfo{pages}{1–13}.
\newblock
\showISBNx{9781450356206}
\urldef\tempurl%
\url{https://doi.org/10.1145/3173574.3173797}
\showURL{%
\tempurl}


\end{thebibliography}

\end{document}